# Effects of high-pressure synthesis on phase formation and superconducting properties of $PrFeAsO_{1-x}F_x$

Priya Singh[1], Konrad Kwatek[2], Tatiana Zajarniuk[3], Tomasz Cetner[1], Jan Mizeracki[1], Shiv J. Singh[1*]

[1]*Institute of High Pressure Physics (IHPP), Polish Academy of Sciences, Sokołowska 29/37, 01-142 Warsaw, Poland*

[2]*Faculty of Physics, Warsaw University of Technology, Koszykowa 75, 00-662, Warsaw, Poland*

[3]*Institute of Physics, Polish Academy of Sciences, Aleja Lotnikow 32/46, 02-668 Warsaw, Poland*

***Corresponding author:**

Email: sjs@unipress.waw.pl

https://orcid.org/0000-0001-5769-1787

# Abstract

Motivated by recent reports of enhanced superconducting performance in several families of iron-based superconductors (IBS) processed by high-pressure (HP) synthesis, we investigate the influence of high gas pressure and high-temperature synthesis (HP-HTS) process on the structural, microstructural, electrical transport, and magnetic properties of Pr-based oxypnictide $PrFeAsO_{1-x}F_x$ (Pr1111) using the processing conditions of 0.5 GPa for 1 h previously optimized for other IBS families. Representative underdoped ($x$ = 0.2), optimal doped ($x$ = 0.3), and overdoped ($x$ = 0.5) compositions from the ambient-pressure electronic phase diagram of Pr1111 are selected to evaluate its composition-dependent effects of HP-HTS. The results demonstrate that HP-HTS enhances fluorine incorporation, improves phase formation, and produces a denser microstructure with improved grain connectivity in the underdoped and optimal doped compositions. Magnetic measurements reveal increased in the superconducting transition temperature ($T_c$) of ~1 K for $x$ = 0.2 and ~6 K for $x$ = 0.3, whereas only a marginal improvement in the critical current density is observed. Electrical resistivity measurements of the underdoped composition show a slight increase in $T_c$ accompanied by a broader resistive transition, indicating residual structural inhomogeneity. In contrast, the overdoped composition exhibits increased impurity phase segregation, accompanied by suppression of superconductivity. These results demonstrate that the effectiveness of HP-HTS in Pr1111 is strongly composition dependent and governed by the interplay among fluorine incorporation, phase stability, and microstructural evolution, highlighting the need for further optimization of the HP-HTS processing conditions.

***Keywords***: Iron-based superconductors, high-pressure synthesis, Critical current density, critical transition temperature.

# I. Introduction

The discovery of superconductivity in $LaFeAsO_{1-x}F_x$ with a transition temperature $T_c$ ~ 26 K by Kamihara et al. in 2008 [1] triggered extensive research into iron-based superconductors (IBS). Owing to their relatively high superconducting transition temperatures ($T_c$), unconventional pairing mechanisms, and the intricate interplay between superconductivity and magnetism, IBS have emerged as one of the most important classes of unconventional superconductors. Among the various IBS families, the rare-earth-based oxypnictides *RE*FeAsO (*RE*1111; *RE* = La, Ce, Pr, Nd, Sm, Gd, etc.), commonly referred to as the 1111 family, exhibit the highest superconducting transition temperatures, reaching up to ~58 K [2], [3] through electron doping by fluorine substitution or co-doping strategies involving thorium and fluorine. Such doping suppresses the antiferromagnetic ground state and stabilizes superconductivity.

Recently, the structural, electronic, magnetic, and thermal properties of $PrFeAsO_{1-x}F_x$ samples synthesized at ambient pressure were systematically investigated, leading to the establishment of a comprehensive electronic phase diagram for the Pr1111 system [4]. This work identified distinct underdoped ($0.15 \leq x < 0.3$), optimal doped ($0.3 \leq x \leq 0.4$), and overdoped ($0.4 < x \leq 1.0$) superconducting regions and demonstrated that $PrFeAsO_{1-x}F_x$ provides an attractive platform for investigating the influence of high-pressure synthesis across the superconducting dome. In IBS, superconductivity originates primarily within the FeAs layers, where subtle modifications of the crystal structure strongly influence the electronic properties. Previous studies have established that structural parameters, such as the As–Fe–As bond angle within the $FeAs_4$ tetrahedron and the pnictogen height ($h_{Pn}$) above the Fe plane play a crucial role in determining the hybridization between Fe 3d and As 4p orbitals, thereby affecting the electronic structure near the Fermi level and the superconducting pairing interactions [5]. In particular, Kuroki *et al.* [5] demonstrated that the pnictogen height strongly modifies the Fermi-surface topology by controlling the appearance or disappearance of characteristic hole Fermi pockets, thereby changing the dominant pairing interaction and potentially switching the superconducting state from a high-$T_c$ nodeless pairing to a low-$T_c$ nodal pairing. Because these structural parameters are highly sensitive to the pressure, the applied pressure has become a powerful tool for tuning the superconducting properties of IBS.

Both externally applied pressure and pressure-assisted synthesis techniques, such as high gas pressure and high-temperature synthesis (HP-HTS), have been widely employed to tune the properties of iron pnictides and chalcogenides [6] [7] [8] [9] [10]. However, these approaches influence superconductivity through fundamentally different mechanisms. Unlike chemical pressure, externally applied physical pressure modifies interatomic distances and local bonding environments without altering the chemical composition [8]. Consequently, measurements performed under applied pressure have revealed significant changes in the superconducting transition temperature, carrier concentration, magnetic ordering, and electronic structure [6]. In contrast, pressure-assisted synthesis applies pressure only during sample preparation, leading to permanent modifications to phase formation, microstructure, and residual lattice strain that remain after pressure release. As a result, HP-HTS often improves grain connectivity, phase purity, and microstructural densification, thereby enhancing the overall quality and performance of polycrystalline superconductors [7].

Among the available high-pressure synthesis methods, HP-HTS based on hot-isostatic pressing (HIP) has attracted considerable attention because it provides a uniform high-pressure gaseous environment, excellent temperature control, and the capability to process relatively large sample volumes [11]. Recent studies have demonstrated the effectiveness of HP-HTS processing in several IBS families under optimized processing conditions [7]. For example, high-pressure growth of $CaKFe_4As_4$ (1144 family) polycrystalline samples resulted in improved microstructure, enhanced superconducting transition temperature, and higher critical current density $J_c$ [12], [13], [14]. Similarly, HP-HTS-treated $SmFeAsO_{0.8}F_{0.2}$ bulks exhibited significantly improved density and critical current density while maintaining their superconducting transition temperature [15]. More recently, high-pressure synthesis using a cubic-anvil technique has also been applied to fluorine-doped SmFeAsO, resulting in an enhancement of the superconducting transition temperature by approximately 3 K and an increase in the $J_c$ by one to two orders of magnitude [16] [17]. In the 11 family, HP-HTS processing of $FeSe_{0.5}Te_{0.5}$ enhanced phase formation, increased $T_c$ by approximately 2–3 K, and improved $J_c$ by nearly an order of magnitude [18], [19], [20]. Collectively, these studies indicate that processing conditions of approximately 0.5 GPa for 1 h have been successfully applied to obtain dense, high-quality superconducting bulks with improved performance across several IBS families [7]. In contrast, the effects of high-pressure synthesis on the $PrFeAsO_{1-x}F_x$ system remain largely unexplored [21]. Although high-pressure growth of PrFeAs(O,F) single crystals has been reported using diamond-anvil-cell techniques for a few fluorine compositions, the obtained crystals were extremely small, limiting comprehensive investigations [22]. Moreover, a previous study on high-pressure synthesis of $PrFeAsO_{0.89}F_{0.11}$ reported an enhancement of $T_c$ to approximately 52 K under high synthesis temperatures [23], suggesting that pressure-assisted growth may provide an effective route for optimizing superconductivity in this system. These studies therefore motivate a systematic investigation of pressure-assisted synthesis across a broader fluorine concentration range in $PrFeAsO_{1-x}F_x$ from under to over doped regions.

Motivated by these considerations, the present work investigates the influence of the HP-HTS process on the representative compositions from the underdoped, optimal doped, and overdoped regions of the $PrFeAsO_{1-x}F_x$ phase diagram. Specifically, samples with nominal fluorine contents $x$ = 0.2, 0.3, and 0.5 are synthesized using both the conventional synthesis method at ambient pressure (CSP-AP) and the HP-HTS technique. The HP-HTS synthesis conditions are selected based on the processing parameters (0.5 GPa, 1 h) previously established to be effective for $CaKFe_4As_4$ (1144) [12], Sm-based 1111 (SmFeAs(O,F)) [15], and 11 (Fe(Se,Te)) [18] families, allowing a direct assessment of whether similar processing conditions are applicable to the Pr1111 system. Comprehensive structural, microstructural, electrical resistivity, and magnetic measurements are performed to elucidate the effects of high-pressure synthesis. Particular emphasis is placed on understanding how HP-HTS influences phase formation, microstructural densification, superconducting homogeneity, and the composition-dependent evolution of the superconducting properties across the underdoped, optimal doped, and overdoped regimes of $PrFeAsO_{1-x}F_x$.

## II. Experimental methods

Polycrystalline $PrFeAsO_{1-x}F_x$ samples with nominal fluorine concentrations $x$ = 0.2, 0.3 and 0.5 were initially synthesized at ambient pressure using CSP-AP method, as described elsewhere [4, 24]. All handling procedures, including weighing, mixing of precursors, regrinding, and pelletization, were performed inside a high-purity argon-filled glove box with oxygen and moisture levels below 1 ppm. To investigate the effects of high-pressure synthesis, the CSP-AP processed samples were subsequently subjected to HP-HTS method [11] as an *ex-situ* process. The quartz ampoules containing the CSP-AP samples were opened inside the glove box, and the obtained pellets were sealed in tantalum (Ta) tubes under an argon atmosphere using an Arc-melting system. The sealed Ta tubes were then placed in the HP-HTS chamber and processed under optimized conditions of 0.5 GPa and 950 °C for 1 h, following the procedure reported previously [7], [4]. For clarity, the samples subjected to HP-HTS are denoted as $x$ = 0.2_HIP, 0.3_HIP, and 0.5_HIP, while the corresponding ambient-pressure samples are denoted simply as $x = 0.2$, 0.3, and 0.5, respectively. Throughout this manuscript, the terms underdoped, optimal doped, and overdoped refer to the nominal fluorine compositions used during sample synthesis. The corresponding experimentally determined fluorine concentrations are presented in Figure 6(a).

Phase formation and crystal structure were investigated by powder X-ray diffraction (XRD) using a Panalytical Empyrean diffractometer with CuKα radiation (λ = 1.5418 Å), operating at 40 kV and 35 mA. Diffraction patterns were collected over the $2\theta$ range of 10°–90° with a step size of 0.013° and a counting time of 300 s per step. Phase identification and structural analysis were carried out using the PDF4+2026 database provided by the International Centre for Diffraction Data (ICDD). Microstructural characterization was performed using a Zeiss Ultra Plus field-emission scanning electron microscope (FE-SEM) equipped with an energy-dispersive X-ray spectroscopy (EDX) system and an ultra-fast detector. Magnetic measurements were conducted using a vibrating sample magnetometer (VSM) integrated with a Quantum Design Physical Property Measurement System (PPMS). The temperature dependence of magnetic susceptibility was measured under both zero-field-cooled (ZFC) and field-cooled (FC) conditions in applied magnetic fields ranging from 20 to 200 Oe over the temperature interval 5–60 K. Quantitative estimates of the superconducting shielding fraction were not performed because the HP-HTS samples possessed small masses and irregular geometries, which introduced significant uncertainties in the demagnetization corrections. Therefore, the magnetic measurements are discussed primarily in terms of relative changes in the superconducting response and transition characteristics. Magnetic hysteresis loops (*M-H*) were recorded at 5 K in magnetic fields up to 9 T. Electrical resistivity measurements were performed using a standard four-probe technique on rectangular specimens cut from the bulk samples. Electrical contacts were prepared using silver paste and copper wires. The temperature dependence of resistivity in zero magnetic field was measured using a closed-cycle refrigerator (CCR).

## III. Results and Discussions

### (i) Structural Analysis

The powder X-ray diffraction (XRD) patterns of $PrFeAsO_{1-x}F_x$ samples synthesized by CSP-AP and by HP-HTS methods are presented in Fig. 1(a–c). The diffraction peaks of all samples can be indexed predominantly to the tetragonal ZrCuSiAs-type structure (space group (P4/nmm)), confirming the formation of the PrFeAs(O,F) superconducting phase, consistent with previous reports on *RE*1111 compounds [25, 16, 4]. In addition to the main phase, weak reflections associated with secondary phases such as PrOF, PrAs, and FeAs/$Fe_2As$ are observed, with their relative abundance depending on both fluorine concentration and synthesis conditions. For the underdoped composition, the HP-HTS sample i.e. $x$ = 0.2_HIP exhibits a noticeable reduction in the PrOF impurity reflections compared with the ambient-pressure sample $x$ = 0.2 as shown in Fig. 1(a). The PrAs impurity phase remains detectable; however, its diffraction intensity is also reduced after HP-HTS processing. These observations indicate improved phase stability and suggest that high-pressure processing promotes more efficient fluorine incorporation into the PrFeAs(O,F) lattice, consistent with the increased fluorine content determined by EDX measurements (Fig. 6(a)). This trend is also consistent with previous studies on $PrFeAsO_{1-x}F_x$ and related *RE*1111 compounds, where fluorine substitution at the oxygen site generally leads to a contraction of the in-plane lattice parameter $a$ and the unit-cell volume $V$ because the ionic radius of $F^-$ is smaller than that of $O^{2-}$ [2, 21, 4]. A similar behavior is observed for the optimal doped composition, as shown in Fig. 1(b). The HP-HTS processed sample $x$ = 0.3_HIP exhibits further suppression of secondary phases, while weak FeAs-related reflections observed in the ambient-pressure sample $x$ = 0.3 become undetectable within the resolution limit of the XRD measurement. The reduction of PrOF impurity phases together with the increased fluorine content obtained from EDX analysis indicates that HP-HTS facilitates fluorine incorporation into the PrFeAs(O,F) lattice and improves phase formation in the near-optimal fluorine doping regime. Such behavior is consistent with the beneficial effects of high-pressure processing reported for other IBS families [18, 7]. On the other hand, the overdoped composition exhibits a different response after HP-HTS treatment. As shown in Fig. 1(c), the $x$ = 0.5_HIP sample displays increased relative intensities of PrOF and PrAs impurity reflections compared with the ambient-pressure sample. This indicates enhanced phase segregation under high-pressure conditions. Although the EDX analysis indicates an increase in the fluorine content after HP-HTS processing, the simultaneous increase in the PrOF and PrAs impurity phases indicates that the additional fluorine is not fully incorporated into the superconducting Pr1111 phase. Instead, once the fluorine solubility limit is exceeded, excess fluorine promotes the formation of competing secondary phases, resulting in enhanced phase segregation. Similar fluorine solubility limitations have been reported in other 1111-type superconductors [21] [2, 16]. Rietveld refinements were carried out for all samples to extract the structural parameters. A representative refinement profile for the $x$ =

0.2_HIP sample is shown in Fig. 1(d), while additional refinement results for the $x = 0.3$_HIP and $x = 0.5$_HIP samples are presented in Supplementary Fig. S1(a) and S1(b), respectively. The Rietveld refinement results for the corresponding ambient-pressure samples ($x = 0.2$, 0.3, and 0.5) have been reported in our previous work [4, 24].

The evolution of the lattice parameters of $PrFeAsO_{1-x}F_x$ synthesized under ambient-pressure and high-pressure conditions is summarized in Fig. 2(a–c). The in-plane lattice parameter *a* and the unit-cell volume *V* exhibit a slight reduction following HP-HTS treatment, indicating a slight overall lattice contraction. This behavior is consistent with the increased fluorine incorporation revealed by the EDX analysis, as increased fluorine substitution at the oxygen site modifies the local bonding environment and consequently the lattice dimensions. The observed lattice evolution is therefore in good agreement with previous reports on Pr1111 and related *RE*1111 superconductors, supporting the conclusion that HP-HTS enhances fluorine incorporation at low and intermediate fluorine concentrations [2, 21, 4]. In contrast, the out-of-plane lattice parameter *c* exhibits a composition-dependent response after HP-HTS treatment, as shown in Fig. 2(b). While the underdoped sample $x = 0.2$_HIP exhibits a slightly reduced *c*-axis parameter compared with the ambient-pressure sample $x = 0.2$, both the optimal doped $x = 0.3$_HIP and overdoped $x = 0.5$_HIP samples show a small expansion along the *c*-axis after HP-HTS treatment. This anisotropic structural response suggests that the high-pressure processing introduces residual lattice modifications that persist after pressure release rather than producing reversible lattice compression. Such changes may originate from residual lattice strain introduced during HP-HTS, variations in defect concentration, and redistribution of fluorine within the crystal structure [21]. The contrasting evolution of the *a* and *c* lattice parameters demonstrates that the structural response to HP-HTS is anisotropic rather than a simple uniform lattice compression. Instead, high-pressure processing modifies the structural state of the PrFeAs(O,F) phase in a composition-dependent manner. In particular, the increase in the *c*-axis parameter observed for the overdoped composition coincides with the enhanced secondary-phase formation, suggesting that excessive fluorine concentrations promote structural instability and phase segregation under HP-HTS conditions. Overall, the structural analysis demonstrates that HP-HTS systematically enhances fluorine incorporation and phase formation in the underdoped and near-optimal compositions, whereas excessive fluorine concentrations in the overdoped regime exceed the fluorine solubility limit of the Pr1111 phase, thereby promoting phase segregation and reducing structural stability of the Pr1111 phase.

These structural observations provide a consistent structural basis for understanding the composition-dependent evolution of the superconducting properties discussed in Section IV.

### (ii) Microstructural Analysis

Representative SEM micrographs of the CSP-AP and HP-HTS processed $PrFeAsO_{1-x}F_x$ samples are presented in Fig. 3(a–f). The ambient-pressure samples $x = 0.2$, 0.3, and 0.5 exhibit relatively porous microstructures characterized by numerous micro- and nanoscale voids together with visible secondary-phase regions, as shown in Fig. 3(b), (d), and (f). Such microstructural features are commonly observed in polycrystalline 1111-type IBS synthesized by the CSP-AP routes [21]. In contrast, the HP-HTS processed samples $x$ = 0.2_HIP, 0.3_HIP, and 0.5_HIP exhibit significantly improved densification, with a marked reduction in micro- and nanoscale porosity and improved grain-to-grain connectivity, as shown in Fig. 3(a), (c), and (e). These observations are consistent with previous studies demonstrating that the HP-HTS process promotes microstructural densification through the simultaneous application of elevated temperature and isostatic pressure [7], [26]. The reduction in porosity and improved grain-to-grain connectivity are expected to facilitate current transport across grain boundaries and mitigate weak-link effects that commonly limit the performance of polycrystalline IBS [27]. The observed microstructural evolution is fully consistent with the XRD and EDX analyses. For the underdoped and optimal doped compositions ($x$ = 0.2_HIP and $x$ = 0.3_HIP), the improved densification is accompanied by a reduction in secondary phases and enhanced fluorine incorporation, indicating that HP-HTS simultaneously promotes phase formation and microstructural refinement. In contrast, although the overdoped sample ($x$ = 0.5_HIP) also exhibits improved densification, XRD analysis reveals increased PrOF and PrAs impurity phases, indicating that microstructural densification alone is insufficient to suppress phase segregation once the fluorine solubility limit is exceeded. Overall, these observations demonstrate that HP-HTS produces a denser and more homogeneous microstructure across all the investigated compositions. However, the superconducting response depends not only on microstructural quality but also on the extent of fluorine incorporation and the resulting phase stability. Consequently, improvements in microstructural densification alone are insufficient to optimize the superconducting properties, particularly in the overdoped regime where excessive fluorine promotes phase segregation, as discussed in the following sections.

### (iii) Electrical Resistivity Measurement

Electrical resistivity measurements were successfully performed for the underdoped HP-HTS processed sample $x$ = 0.2_HIP and compared with its ambient-pressure counterpart $x$ = 0.2 [4]. Reliable resistivity measurements could not be obtained for the $x$ = 0.3_HIP and $x$ = 0.5_HIP samples because the available specimens were not sufficiently large for proper electrical contact preparation. Although HP-HTS is capable of producing large and dense bulk samples [11, 12], achieving this requires optimization of the processing parameters for the specific material system. The present results indicate that further optimization of the HP-HTS conditions is necessary to obtain larger Pr1111 specimens suitable for comprehensive transport characterization. The temperature dependence of the normalized resistivity in the low-temperature region is shown in Fig. 4, while the inset figure presents the normalized resistivity over the extended temperature range up to 300 K. Both samples exhibit metallic behavior in the normal state, characterized by a monotonic decrease in resistivity upon cooling from room temperature. A superconducting transition is observed at low temperatures with an onset transition temperature of $T_c^{onset}$ ~ 47 K for the ambient-pressure sample $x$ = 0.2, consistent with our previous report [4]. The criteria used to determine the onset ($T_c^{onset}$) and offset ($T_c^{offset}$) superconducting transition are illustrated in the Supplementary Figure S2. Following HP-HTS processing, the onset superconducting transition temperature increases by approximately 1 K, indicating a modest enhancement of the superconducting transition temperature in the underdoped Pr1111 composition. This improvement is consistent with the enhanced fluorine incorporation revealed by the EDX analysis (Fig. 6(a)), suggesting that high-pressure processing promotes more effective fluorine substitution into the PrFeAs(O,F) lattice. Another notable difference is also observed in the width of the superconducting transition, defined as ($\Delta T = T_c^{onset} - T_c^{offset}$). The HP-HTS processed sample $x$ = 0.2_HIP exhibits a broader resistive transition compared with the ambient-pressure sample ($x$ = 0.2). This broadening suggests increased local superconducting inhomogeneity despite the improved phase purity, enhanced fluorine incorporation, and improved microstructural densification revealed by the XRD, EDX, and SEM analyses. The broadened transition is most likely associated with residual lattice strain and local compositional variations introduced during HP-HTS process and retained after pressure release. Unlike *in-situ* high-pressure experiments, where non-hydrostatic pressure conditions broaden the superconducting transition during measurement [28], the present effect is attributed to permanent structural modifications induced during high-pressure synthesis. The composition-dependent changes in the lattice parameter $c$ following HP-HTS processing further support the presence of such residual structural alterations. These structural modifications may influence the local electronic environment of the superconducting FeAs layers, giving rise to spatial variations in superconducting properties. Consequently, the superconducting transition proceeds through a broader percolative process, leading to an increased transition width despite the slight enhancement of the $T_c^{onset}$. Overall, these results demonstrate that HP-HTS under the present processing conditions (0.5 GPa, 1 h) improves phase formation, fluorine incorporation, and microstructural densification in the underdoped Pr1111 composition. The modest increase in the superconducting transition temperature is accompanied by a broader resistive transition, suggesting that the beneficial effect of enhanced fluorine incorporation coexists with residual structural inhomogeneity introduced during HP-HTS processing. These observations are further supported by the magnetic measurements presented in the following section, which provide additional evidence that HP-HTS modifies both the superconducting transition temperature and the superconducting homogeneity.

## (iv) Magnetic Properties

The magnetic properties of HP-HTS processed $PrFeAsO_{1-x}F_x$ samples representing the underdoped $x$ = 0.2_HP, optimal doped $x$ = 0.3_HIP, and overdoped $x$ = 0.5_HIP compositions are investigated by zero-field-cooled (ZFC) and field-cooled (FC) magnetization measurements in the temperature range of 5–60 K. The results are compared with those of the corresponding CSP-AP processed Pr1111 samples reported previously [4]. Measurements for the $x$ = 0.2_HIP and $x$ = 0.3_HIP samples were performed under an applied magnetic field of 20 Oe, whereas a field of 100 Oe was used for the $x$ = 0.5_HIP sample because of its weaker superconducting response. The superconducting transition temperature $T_c^{mag}$ was determined from the onset of the diamagnetic transition in the ZFC magnetization curve. The corresponding ZFC and FC magnetization curves are presented in Fig 5(a–c). For the underdoped composition ($x$ = 0.2_HIP), the superconducting transition temperature ($T_c^{mag}$) determined from the magnetic measurements exhibits a modest enhancement of approximately 1 K compared with the corresponding CSP-AP sample ($T_c^{mag}$ ~46.5 K). A much more pronounced enhancement is observed for the optimal doped composition, where the superconducting transition temperature of the $x$ = 0.3_HIP sample increases by ~6 K after HP-HTS processing. These improvements are fully consistent with the enhanced fluorine incorporation revealed by the EDX analysis (Fig. 6(a)), indicating that HP-HTS promotes more effective fluorine substitution into the PrFeAs(O,F) lattice and shifts the carrier concentration toward the optimum superconducting regime. In addition to the enhancement of $T_c$, the HP-HTS processed $x$ = 0.2_HIP sample, shown in Fig. 5(a), exhibits a stronger diamagnetic response below the superconducting transition than the corresponding CSP-AP sample, indicating improved superconducting connectivity and reduced weak-link effects. A similar improvement is observed for the optimal doped composition (Fig. 5(b)), where the $x = 0.3$_HIP sample displays both an enhanced diamagnetic response and a reduced separation between the ZFC and FC magnetization curves compared with the ambient-pressure sample. Such behavior is commonly associated with improved intergranular current transport and reduced weak-link effects in polycrystalline IBS [27]. These magnetic observations are fully consistent with the SEM results, which demonstrate improved microstructural densification and enhanced grain-to-grain connectivity following HP-HTS processing. A markedly different behavior is observed for the overdoped sample $x = 0.5$_HIP. As shown in Fig. 5(c), the superconducting transition temperature decreases by approximately 2 K compared with the corresponding CSP-AP sample ($T_c^{mag}$ ~49 K) [4]. In addition, the sample exhibits a weaker diamagnetic response together with enhanced separation between the ZFC and FC magnetization curves, indicating degradation of the superconducting state and stronger weak-link behavior. These observations correlate well with the structural analysis, which revealed increased concentrations of PrOF and PrAs impurity phases after HP-HTS treatment. The results indicate that once the fluorine solubility limit of the Pr1111 phase is exceeded, additional fluorine promotes phase segregation rather than further incorporation into the superconducting phase. Consequently, the detrimental effects of structural instability and secondary-phase formation outweigh the beneficial effects arising from improved microstructural densification.

The suppression of superconductivity in the overdoped HP-HTS sample is also qualitatively consistent with previous pressure studies on 1111-type iron-based superconductors, where excessive carrier concentration combined with pressure-induced structural modifications resulted in a reduction of the superconducting transition temperature [28]. Therefore, the magnetic measurements further demonstrate that the influence of HP-HTS on $PrFeAsO_{1-x}F_x$ is strongly composition dependent. HP-HTS enhances the superconducting transition temperature and improves superconducting connectivity in the underdoped and optimal doped regimes through enhanced fluorine incorporation and improved phase formation, and microstructural refinement. In contrast, once the fluorine solubility limit is exceeded, enhanced phase segregation suppresses superconductivity despite the improved microstructural densification. These magnetic observations are fully consistent with the structural and microstructural analyses, which demonstrate improved phase formation, microstructural densification, and enhanced grain-to-grain connectivity following HP-HTS processing.

To further evaluate the influence of HP-HTS process on the superconducting current-carrying capability, magnetic hysteresis loop (*M-H*) measurements are performed for the $x$ = 0.2_HIP sample, for which a well-defined specimen geometry was available. Similar measurements could not be carried out for the $x$ = 0.3_HIP and $x$ = 0.5_HIP samples because of their small size and irregular shapes, which prevented reliable determination of the sample dimensions required for critical current density calculations. The magnetic critical current density $J_c$ of the CSP-AP processed sample $x = 0.2$ and HP-HTS processed sample $x$ = 0.2_HIP samples is estimated using the Bean critical-state model [29]: $J_c = \frac{20\Delta m}{Va\left(1-\frac{a}{3b}\right)}$ where $\Delta m$ is the width of the magnetic hysteresis loop measured from the difference between the magnetization values recorded during increasing and decreasing magnetic-field sweeps, $a$ and $b$ ($a<b$) are the sample dimensions perpendicular to the applied magnetic field, and $V$ is the sample volume. The field dependence of $J_c$ is shown in Fig. 5(d). The HP-HTS sample $x$ = 0.2_HIP exhibits only a marginal enhancement of the critical current density compared with the corresponding CSP-AP sample $x = 0.2$. Although the XRD, SEM, and EDX analyses demonstrated improved phase formation, enhanced fluorine incorporation, reduced porosity, and better grain-to-grain connectivity following HP-HTS processing, these improvements do not translate into a substantial increase in the current-carrying capability. This observation suggests that microstructural densification alone is insufficient to significantly enhance $J_c$ in the Pr1111 system and that other factors, such as the effectiveness of flux pinning and intergranular current transport, remain important limitations. It is noteworthy that the response of $PrFeAsO_{0.8}F_{0.2}$ to HP-HTS process differs from that reported for several other IBS processed under similar HP-HTS conditions (0.5 GPa, 1 h), including $CaKFe_4As_4$ [12], [13], [14], $SmFeAsO_{0.8}F_{0.2}$ [15] and $FeSe_{0.5}Te_{0.5}$ [18], [19], [20], where pronounced improvements in critical current density and superconducting performance were reported. In contrast, the relatively modest improvement observed for $PrFeAsO_{0.8}F_{0.2}$ indicates that the effectiveness of HP-HTS is strongly material dependent. For the Pr1111 system, enhanced densification and improved fluorine incorporation alone are insufficient to produce a substantial increase in $J_c$, suggesting that flux pinning, grain-boundary properties, and other microstructural and superconducting factors also influence the current-carrying capability. These results are consistent with the magnetic and transport measurements presented above and further support the overall conclusion that the influence of HP-HTS depends on both the iron-based superconductor family and its electronic doping regime.

## IV. Discussion

To assess the effectiveness of fluorine incorporation during synthesis, energy-dispersive X-ray spectroscopy (EDX) measurements were performed for all these samples. A comparison between the nominal fluorine concentration ($x$) and the experimentally determined fluorine content ($x_{act}$) is presented in Fig. 6(a). For the ambient-pressure sample with nominal composition $x = 0.2$, the measured fluorine content is approximately $x_{act} \approx 0.12$, whereas the HP-HTS processed sample ($x$ = 0.2_HIP) exhibits an increased fluorine content of $x_{act} \approx 0.19$. Similarly, for the nominal composition $x = 0.3$, the fluorine content increases from $x_{act} \approx 0.20$ for the CSP-AP sample to $x_{act} \approx 0.25$ after HP-HTS treatment. These increases are accompanied by a reduction in the PrOF impurity phase observed in the XRD analysis, demonstrating that HP-HTS enhances fluorine incorporation into the PrFeAs(O,F) lattice at low and intermediate fluorine concentrations. For the nominal composition $x = 0.5$, the measured fluorine concentration also increases after HP-HTS processing, reaching $x_{act} \approx 0.38$. However, unlike the lower fluorine compositions, the XRD patterns reveal increased concentrations of PrOF and PrAs impurity phases. Although the EDX analysis indicates a higher fluorine content, the simultaneous increase in secondary phases demonstrates that the additional fluorine is not effectively incorporated into the superconducting Pr1111 phase but instead promotes phase segregation. The apparent discrepancy between the EDX and XRD results may partly arise from the larger uncertainty of EDX measurements associated with the increased amount of secondary phases, compositional inhomogeneity, and the limited size and rough surface of the analyzed specimen. Nevertheless, the XRD results clearly indicate that, under the present HP-HTS conditions, excessive fluorine concentrations exceed the fluorine solubility limit of the Pr1111 phase, resulting in the formation of competing impurity phases.

Figure 6(b) summarizes the superconducting transition temperatures determined from electrical resistivity ($T_c^{onset}$) and magnetic susceptibility ($T_c^{mag}$) as a function of the nominal fluorine concentration ($x$). The HP-HTS processed sample $x$ = 0.2_HIP exhibits a modest increase in the superconducting transition of approximately 1 K compared with the corresponding CSP-AP sample ($x = 0.2$). A much more pronounced enhancement is observed for the optimal doped composition, where $T_c^{mag}$ of the $x$ = 0.3_HIP sample increases by approximately 6 K relative to the ambient-pressure synthesized sample ($x$ = 0.3). In contrast, the overdoped composition ($x$ = 0.5_HIP) exhibits a reduction in $T_c^{mag}$ of approximately 2 K related to the corresponding CSP-AP sample $x$ = 0.5. The evolution of $T_c^{mag}$ closely follows the composition-dependent changes in fluorine incorporation revealed by the EDX analysis (Figure 6(a)). However, the relationship is clearly non-monotonic. Enhanced fluorine incorporation improves the superconductivity in the underdoped and optimal doped compositions by shifting the carrier concentration closer to the optimum superconducting regime. In contrast, further fluorine incorporation in the overdoped regime suppresses the superconductivity because the fluorine solubility limit is exceeded, leading to phase segregation and the formation of secondary phases [4].

Overall, the present results demonstrate that the influence of HP-HTS on $PrFeAsO_{1-x}F_x$ is strongly composition dependent and differs from the more uniformly beneficial effects reported for several other IBS families processed under comparable high-pressure conditions [11, 7, 21] [18, 14]. Structural and microstructural analyses reveal that HP-HTS improves phase formation, fluorine incorporation, and microstructural densification, particularly in the underdoped and optimal doped regimes. Nevertheless, only a marginal enhancement of the critical current density ($J_c$) is observed for the $x$ = 0.2_HIP sample, indicating that enhanced fluorine incorporation primarily optimizes the superconducting carrier concentration, whereas the current-carrying capability remains governed by additional factors such as flux pinning and grain-boundary connectivity, including weak-link behavior at grain boundaries. The present results further indicate that HP-HTS influences the superconductivity through two concurrent effects. The first is enhanced fluorine incorporation, which shifts the carrier concentration toward the optimum superconducting regime and increases $T_c$. The second is the introduction of residual lattice strain and local structural inhomogeneity during high-pressure synthesis, which broadens the superconducting transition despite improvements in phase purity and microstructural connectivity. These effects are particularly evident for the underdoped composition, where a broader resistive transition is observed after HP-HTS processing. The observed anisotropic lattice evolution following HP-HTS, particularly the composition-dependent variation of the $c$-axis parameter, suggests that high-pressure processing modifies the local structural environment of the FeAs layers. Although the present diffraction data do not permit direct determination of the Fe–As bond lengths, As–Fe–As bond angles, or pnictogen height, previous studies have shown that these structural parameters strongly influence the electronic structure and superconducting properties of *RE*1111 compounds by modifying the Fe 3d–As 4p hybridization and Fermi-surface topology [30, 31]. Therefore, the enhanced fluorine incorporation together with residual lattice strain and possible local fluorine redistribution introduced during HP-HTS may modify the local structural environment of the FeAs layers. The broader superconducting transition observed after HP-HTS indicates that these beneficial and detrimental effects coexist, resulting in only modest improvements in the bulk superconducting properties despite enhanced fluorine incorporation and improved microstructural densification under the present processing conditions.

A markedly different response is observed in the overdoped fluorine doped regime. For the $x$ = 0.5_HIP sample, the increase in PrOF and PrAs impurity phases demonstrates that excessive fluorine destabilizes the superconducting Pr1111 phase during HP-HTS, resulting in enhanced phase segregation and suppression of superconductivity. This observation demonstrates that enhanced fluorine incorporation alone does not guarantee improved superconducting performance; rather, superconductivity is optimized only within a limited fluorine doping window. The behavior of $PrFeAsO_{1-x}F_x$ contrasts with that reported for other IBS processed under similar HP-HTS conditions (0.5 GPa, 1 h), including $CaKFe_4As_4$ (1144) [12], $SmFeAsO_{0.8}F_{0.2}$ (1111) [15], and $FeSe_{0.5}Te_{0.5}$ (11) [18], where comparable HP-HTS conditions produced simultaneous improvements in both $T_c$ and $J_c$. In the Pr1111 system, however, the enhancement of superconducting performance is confined to a limited fluorine concentration range, while excessive fluorine promotes structural instability and phase segregation. Therefore, the present results demonstrate that, for the investigated compositions, HP-HTS does not universally enhance the superconducting properties of $PrFeAsO_{1-x}F_x$. Instead, its effectiveness is governed by the fluorine concentration. For the investigated underdoped and optimal doped compositions, HP-HTS is beneficial, where enhanced fluorine incorporation improves phase formation and increases the superconducting transition temperature. In contrast, once the fluorine solubility limit is exceeded, further fluorine incorporation promotes phase segregation and suppresses superconductivity despite the improved microstructural densification. These findings demonstrate that the optimization window for pressure-assisted synthesis is considerably narrower in the Pr1111 system than in other IBS families. Further systematic studies involving variations in synthesis pressure, temperature, and dwell time are therefore required to establish the optimum processing conditions and determine whether the beneficial effects of HP-HTS can be extended across the entire superconducting phase diagram of the Pr1111 system. Although electrical resistivity measurements were unavailable for the HP-HTS-treated $x$ = 0.3_HIP and $x$ = 0.5_HIP samples because of the limited specimen size, the combined magnetic susceptibility, EDX, and XRD analyses consistently support the observed composition-dependent evolution of the superconducting properties after HP-HTS treatment. The absence of transport data for these two compositions therefore does not alter the overall conclusions of the present work, although future studies using larger optimized HP-HTS specimens will enable a more comprehensive transport characterization.

## V. Conclusion

The effects of HP-HTS under the present processing conditions (0.5 GPa, 1 h) were systematically investigated for representative underdoped ($x$ = 0.2), optimal doped ($x$ = 0.3), and overdoped ($x$ = 0.5) compositions of the $PrFeAsO_{1-x}F_x$ system. The results demonstrate that, for the three representative compositions investigated, the influence of HP-HTS is strongly composition dependent. For the underdoped and optimal doped compositions, HP-HTS enhances fluorine incorporation, improves phase formation, and produces a denser microstructure with better grain-to-grain connectivity. These improvements are accompanied by increases in the superconducting transition temperature of ~1 K for $x$ = 0.2 and 6 K for $x$ = 0.3. Nevertheless, only a marginal enhancement of $J_c$ is observed, indicating that improved phase formation, fluorine incorporation, and microstructural densification alone are insufficient to substantially enhance the current-carrying capability of the Pr1111 system. In addition, the broader resistive superconducting transition observed for the underdoped composition suggests that residual structural inhomogeneity introduced during HP-HTS coexists with the enhanced $T_c$. In contrast, the overdoped composition ($x$ = 0.5) exhibits increased PrOF and PrAs phase segregation after HP-HTS processing, accompanied by suppression of the superconducting response. These observations indicate that once the fluorine solubility limit of the Pr1111 phase is exceeded, additional fluorine promotes secondary-phase formation rather than further incorporation into the superconducting Pr1111 phase, thereby limiting the beneficial effects of improved microstructural densification. Overall, the present study demonstrates that the effectiveness of HP-HTS in $PrFeAsO_{1-x}F_x$ is governed by the interplay among fluorine incorporation, phase stability, and microstructural evolution. Unlike previously reported HP-HTS studies on $CaKFe_4As_4$, $SmFeAsO_{0.8}F_{0.2}$, and $FeSe_{0.5}Te_{0.5}$, the present results indicate that the beneficial effects of HP-HTS are confined to the investigated fluorine doping range under the present processing conditions. These findings highlight the family- and composition-dependent nature of high-pressure synthesis in IBS and indicate that further optimization of the HP-HTS processing parameters, including applied pressure, temperature, and dwell time, is required to fully realize the potential of HP-HTS for the Pr1111 system. In particular, future work will focus on optimizing these processing conditions to produce larger Pr1111 bulk specimens suitable for comprehensive electrical transport measurements over a wider fluorine composition range.

## CRediT authorship contribution statement

**Priya Singh:** Writing – review & editing, Writing – original draft, Investigation, Formal analysis, Data curation. **Konrad Kwatek:** Formal analysis, Investigation, Resources, Data curation, Writing – review & editing. **Tatiana Zajarniuk:** Data curation, Investigation, Resources, Writing – review & editing. **Tomasz Cetner:** Resources, Data curation. **Jan Mizeracki:** Resources, Data curation. **Shiv J. Singh:** Writing–review & editing, Writing – original draft, Visualization, Validation, Supervision, Software, Resources, Methodology, Investigation, Funding acquisition, Formal analysis, Conceptualization.

## Declaration of competing interest

The authors declare that they have no known competing financial interests or personal relationships that could have appeared to influence the work reported in this paper.

## Data availability

The raw/processed data required to reproduce these findings cannot be shared at this time due to technical or time limitations. Data are available upon request to the corresponding author.

**Acknowledgments:**

The work was funded by the SONATA-BIS 11 project (Registration number: 2021/42/E/ST5/00262) and the Weave-UNISONO project (Registration number: 2025/07/Y/ST5/00116) sponsored by National Science Centre (NCN), Poland. SJS acknowledges financial support from National Science Centre (NCN), Poland through research projects number: 2021/42/E/ST5/00262 and 2025/07/Y/ST5/00116.

**Figure 1**: Powder X-ray diffraction (XRD) patterns of $PrFeAsO_{1-x}F_x$ samples synthesized by the CSP-AP and HP-HTS methods for : (**a**) $x$ = 0.2_HIP and $x$ = 0.2, (**b**) $x$ = 0.3_HIP and $x$ = 0.3 and (**c**) $x$ = 0.5_HIP and $x$ = 0.5. The diffraction peaks are indexed to the tetragonal ZrCuSiAs-type structure (space group (P4/nmm)), and the impurity phases are indicated. (**d**) Representative Rietveld refinement for the $x$ = 0.2_HIP sample. Open circles represent the experimental data, the red solid line represents the calculated profile, the blue line at the bottom represents the difference between the experimental and calculated profiles, and the vertical tick marks indicate the Bragg reflection positions of the PrFeAs(O,F), PrOF, and FeAs phases.

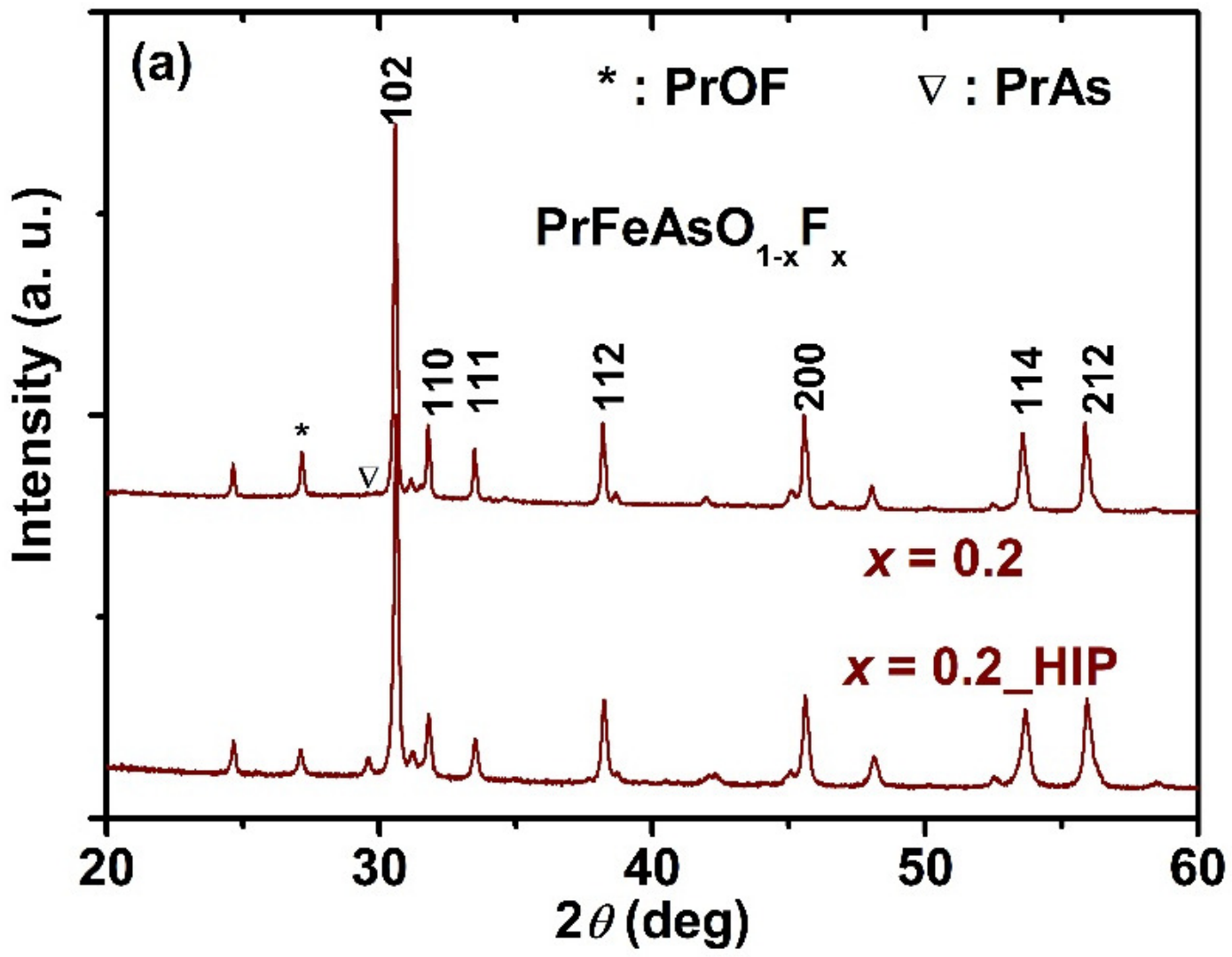

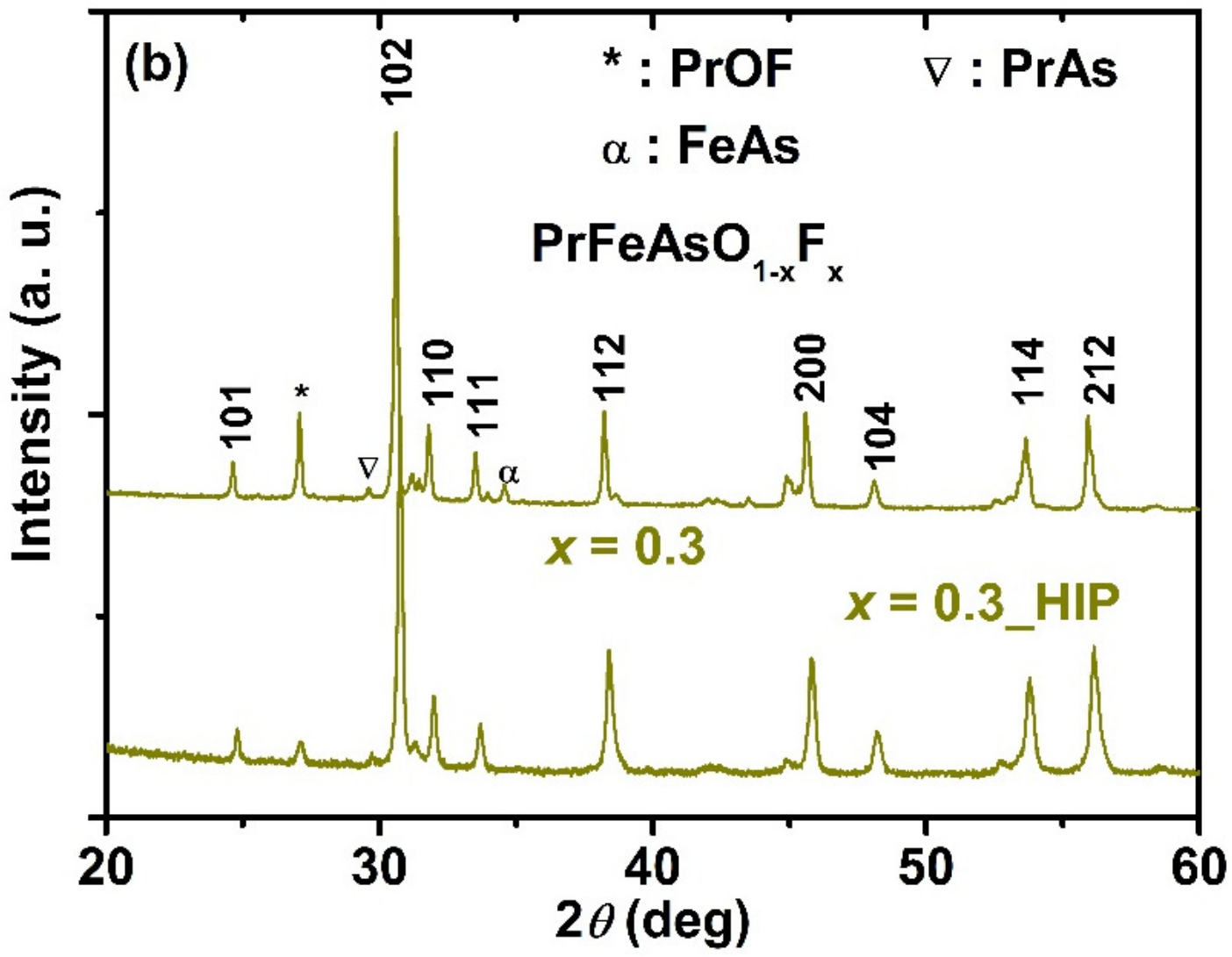

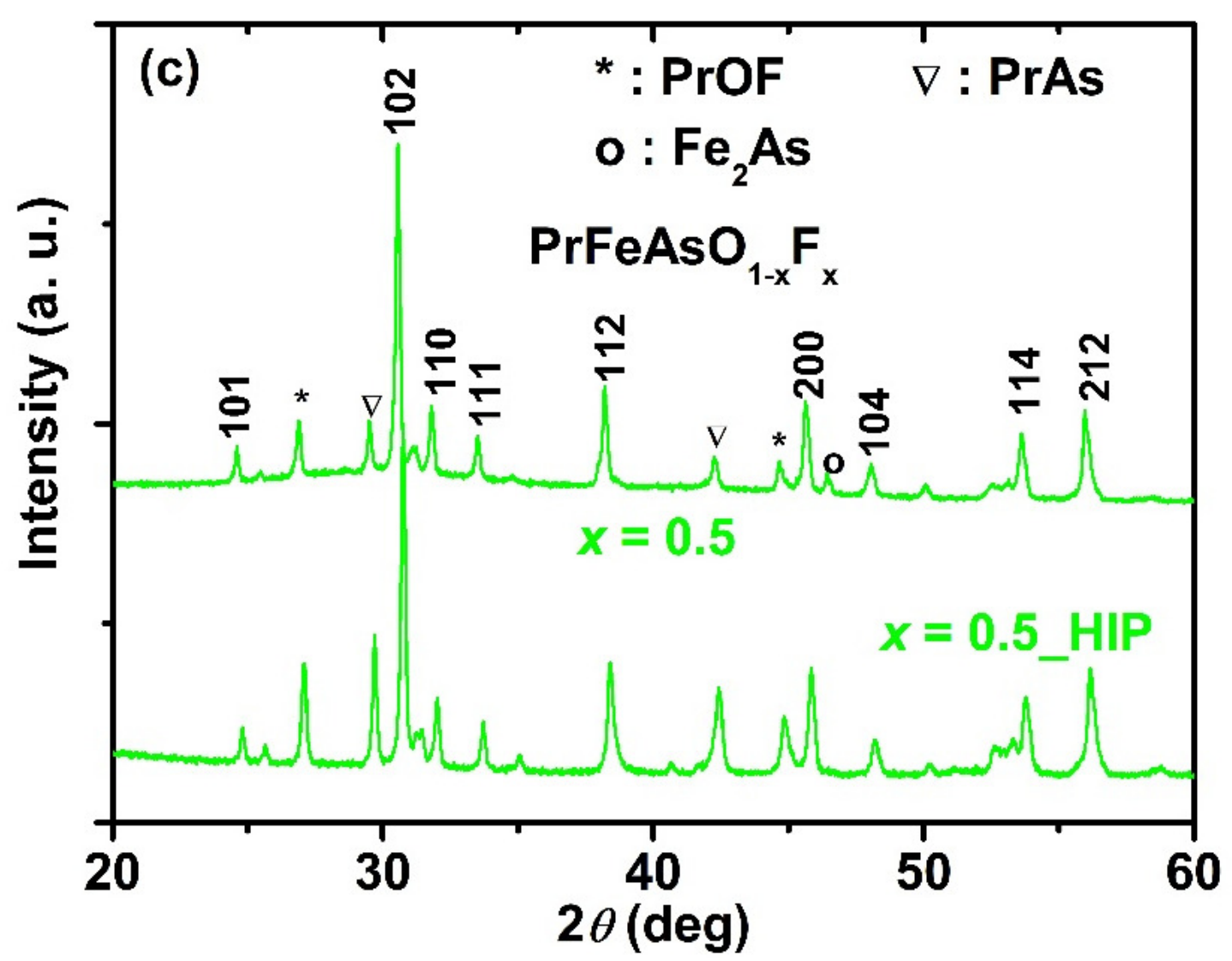
(c)
* : PrOF
∇ : PrAs
o : Fe2As
PrFeAsO1-xFx
101
102
110
111
112
200
104
114
212
x = 0.5
x = 0.5_HIP
Intensity (a. u.)
2θ (deg)
20
30
40
50
60

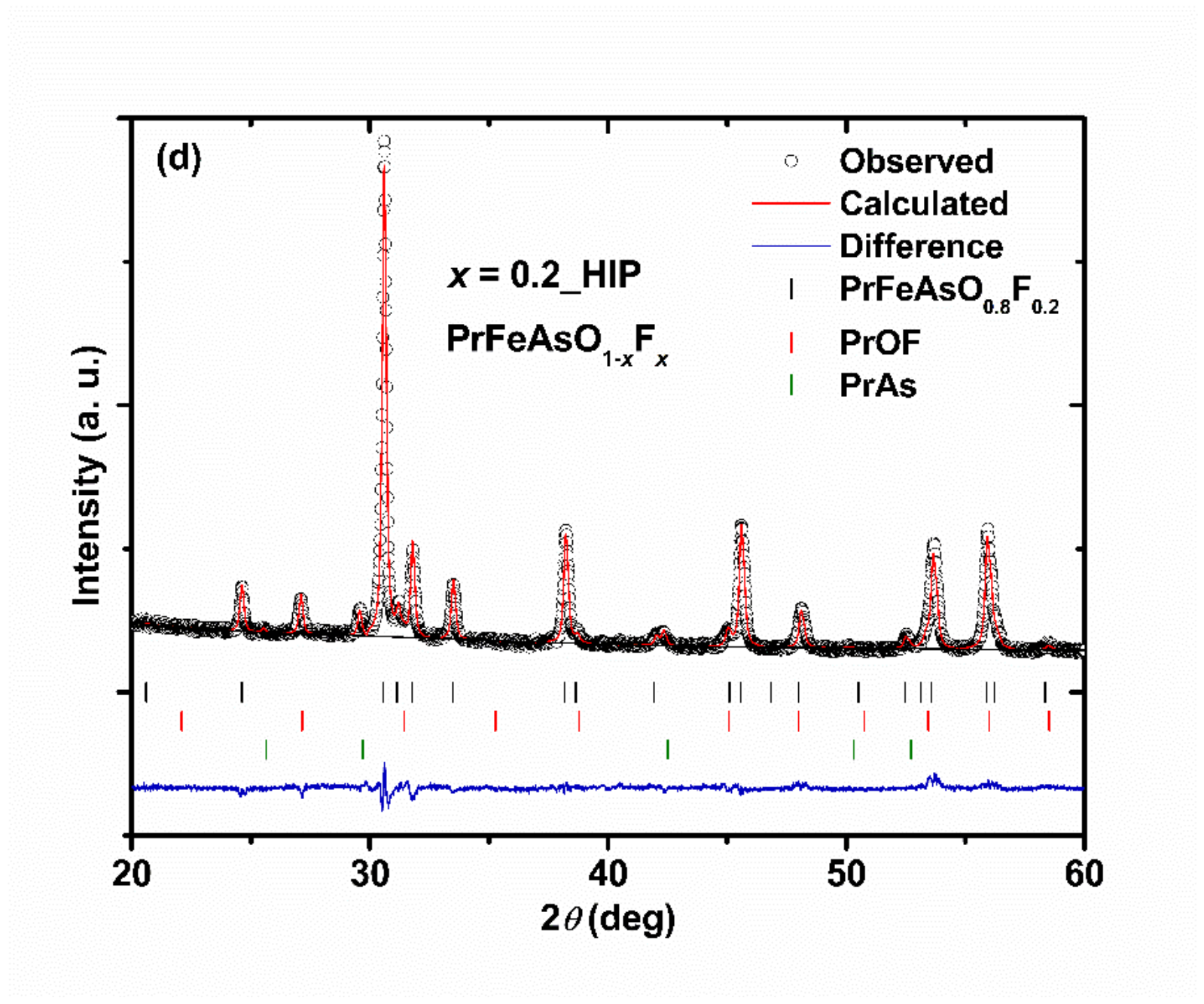
(d)
Observed
Calculated
Difference
PrFeAsO0.8F0.2
PrOF
PrAs
x = 0.2_HIP
PrFeAsO1-xFx
Intensity (a. u.)
2θ (deg)
20
30
40
50
60

**Figure 2**: Variation of the (**a**) lattice parameter *a*, (**b**) lattice parameter *c*, and (**c**) unit-cell volume *V* of $PrFeAsO_{1-x}F_x$ samples synthesized by the CSP-AP and HP-HTS methods as a function of the nominal fluorine concentration (*x*).

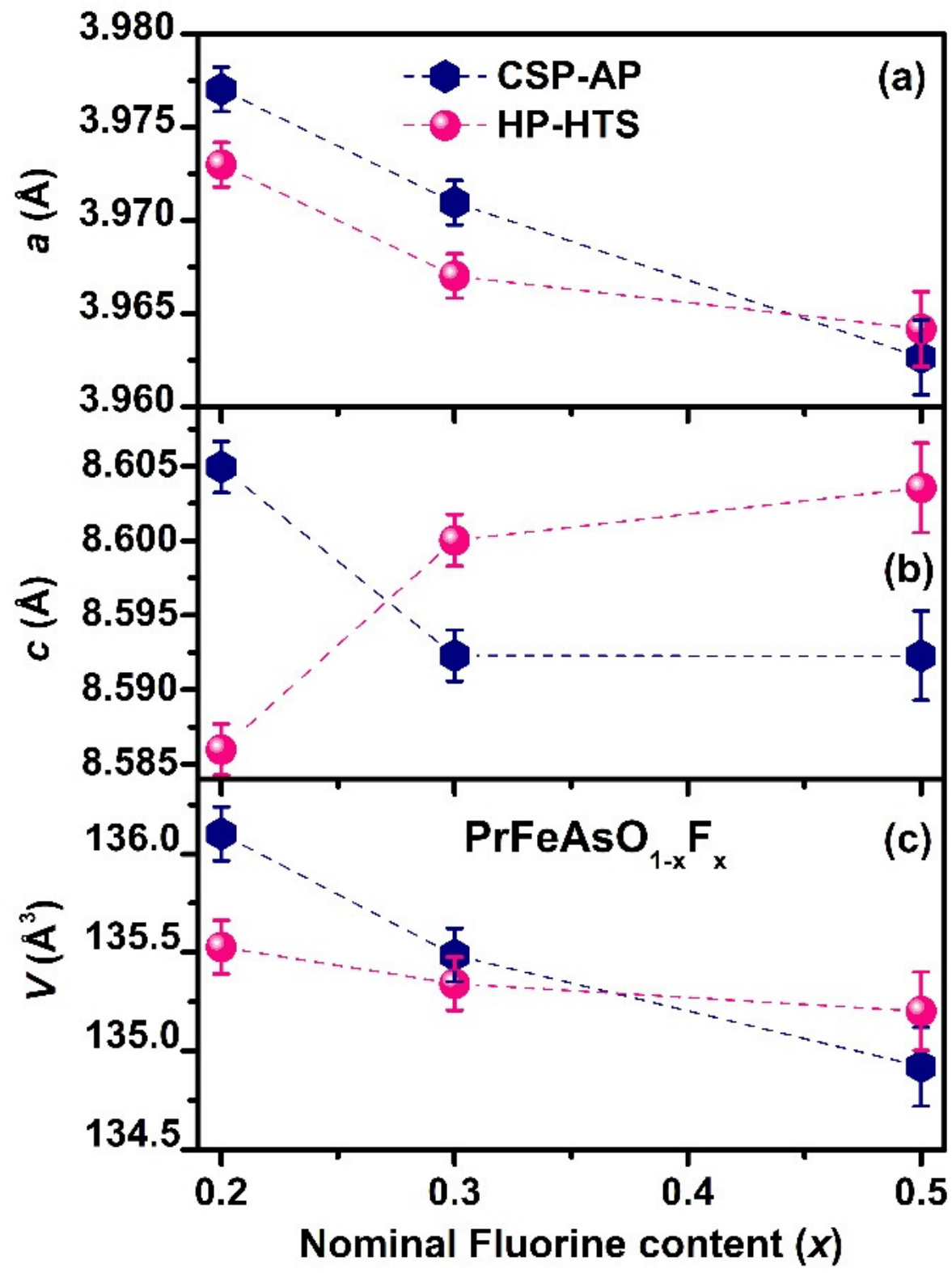

**Figure 3**: Back-scattered electron (BSE) images of $PrFeAsO_{1-x}F_x$ samples synthesized by HP-HTS and CSP-AP methods are depicted in order of increasing fluorine concentration for: (**a**)-(**b**) $x = 0.2$_HIP and $x = 0.2$, (**c**)-(**d**) $x = 0.3$_HIP and $x = 0.3$. (**e**)-(**f**) $x = 0.5$_HIP and $x = 0.5$, respectively.

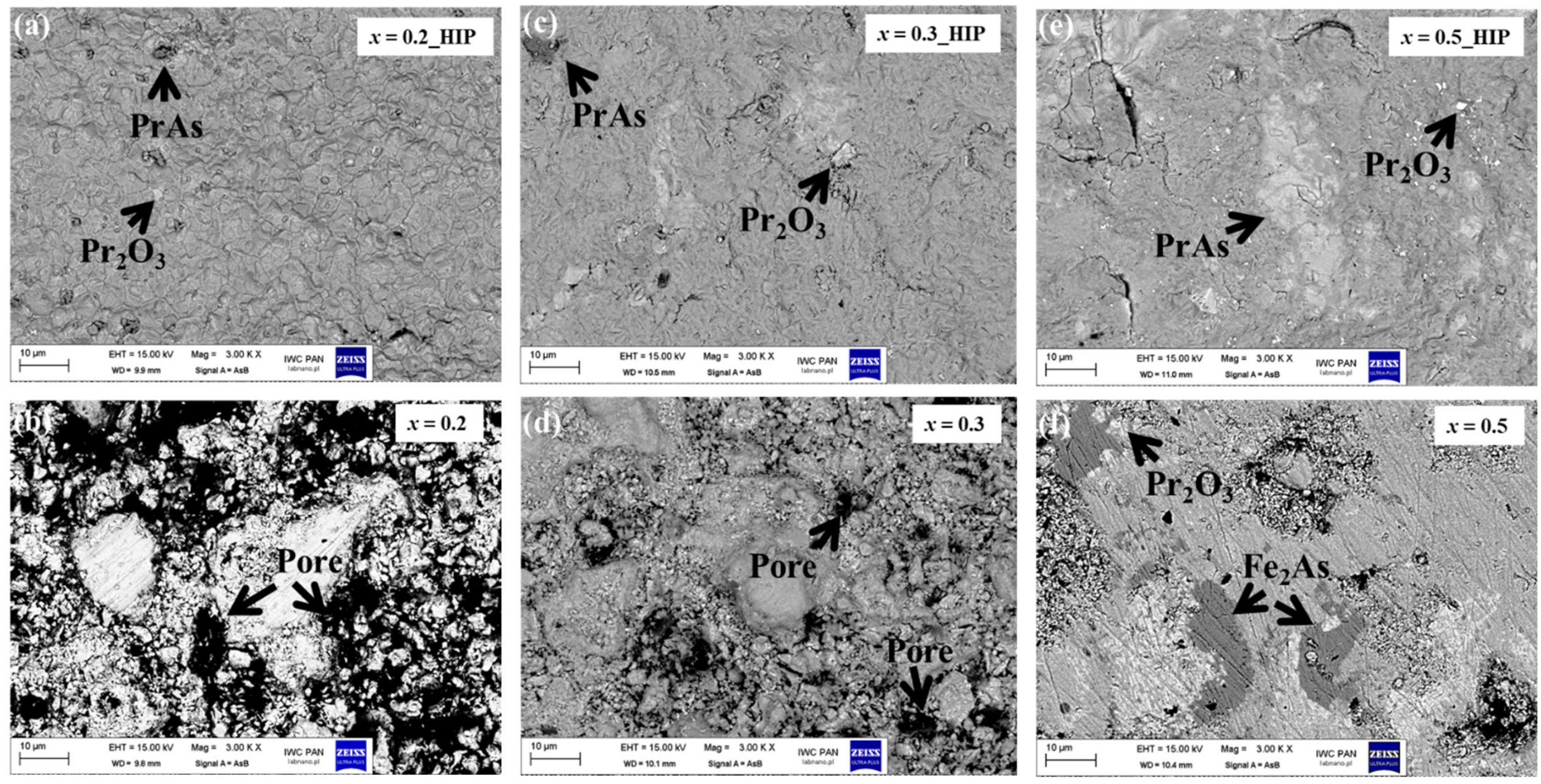

**Figure 4**: Temperature dependence of the normalized electrical resistivity, $\rho/\rho_{60K}$, for the HP-HTS processed sample $x$ = 0.2_HIP and the corresponding CSP-AP processed sample $x$ = 0.2 in the temperature range 7–60 K. The inset figure shows the normalized resistivity, ($\rho/\rho_{300K}$) over the extended temperature range 7–300 K. Reliable resistivity measurements for the $x$ = 0.3_HIP and $x$ = 0.5_HIP samples could not be performed because the available specimens were too small for proper four-probe electrical contact preparation. The definition of the superconducting transition temperatures used in this work is illustrated in Supplementary Fig. S2.

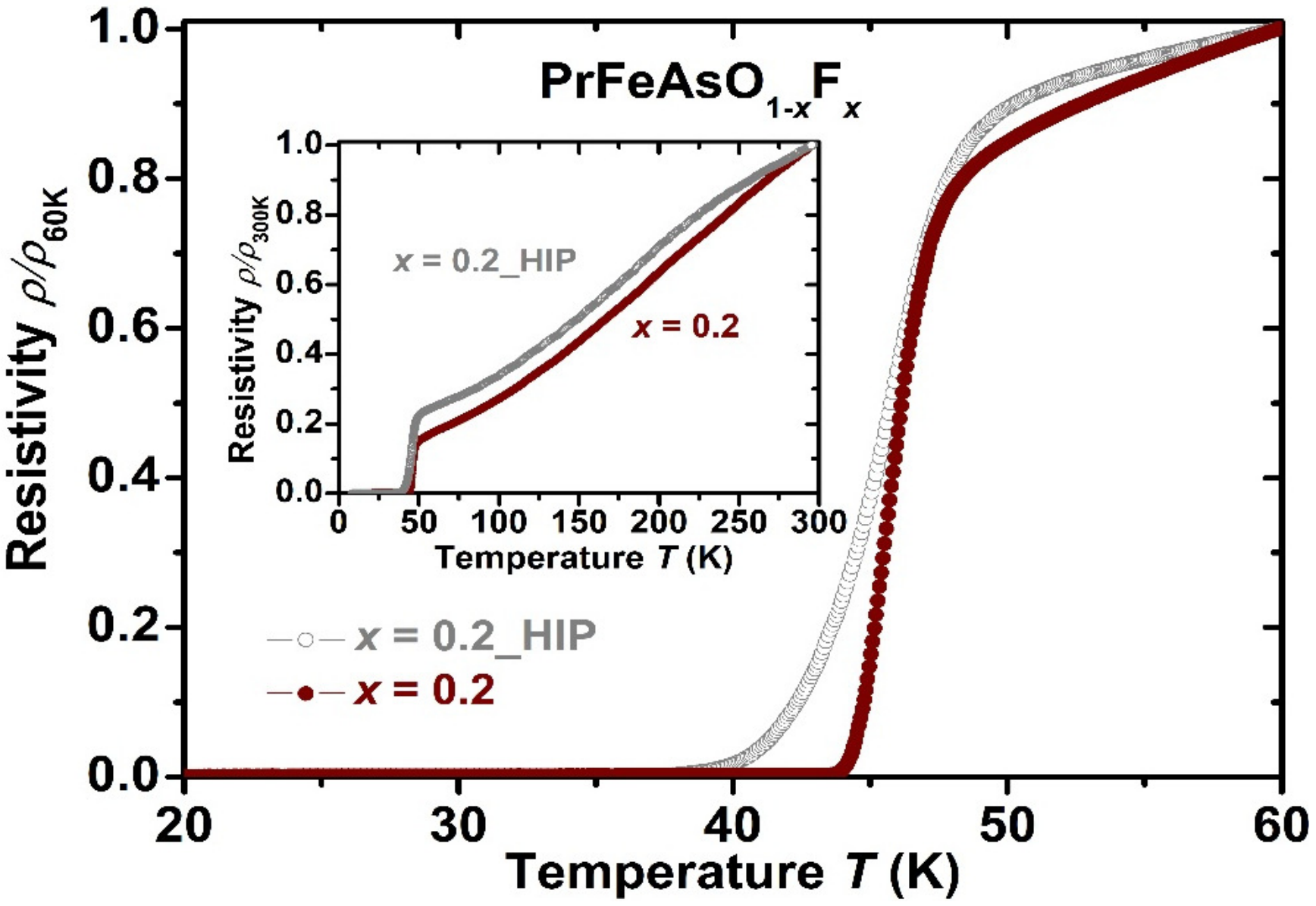

**Figure 5**: Temperature dependence of the normalized magnetization, ($M/M_{5K}$) measured under zero-field-cooled (ZFC) and field-cooled (FC) conditions for $PrFeAsO_{1-x}F_x$ samples synthesized by HP-HTS and CSP-AP methods: (**a**) $x$ = 0.2_HIP and $x$ = 0.2, (**b**) $x$ = 0.3_HIP and $x$ = 0.3 and (**c**) $x$ = 0.5_HIP and $x$ = 0.5. The arrows indicate the superconducting transition temperatures determined from magnetization measurements ($T_c^{mag}$). (**d**) The critical current density, ($J_c$) as a function of applied magnetic field, $\mu_0 H$, for $PrFeAsO_{0.8}F_{0.2}$ synthesized by the CSP-AP ($x$ = 0.2) and HP-HTS methods ($x$ = 0.2_HIP). For comparison, the corresponding $J_c(H)$ data for $CaKFe_4As_4$ [12], $SmFeAsO_{0.8}F_{0.2}$ [15], $FeSe_{0.5}Te_{0.5}$ [18] are also included in Figure 5(d).

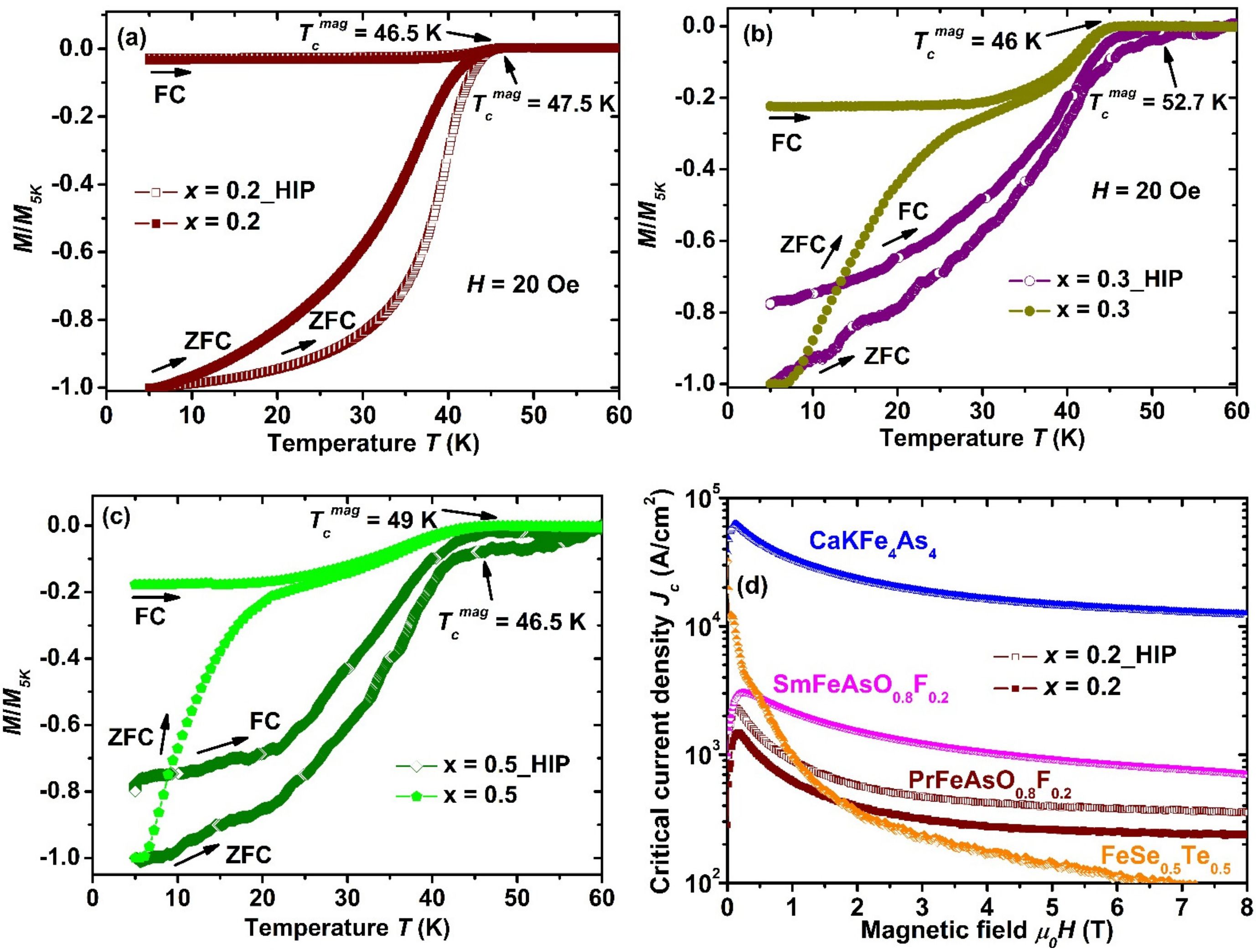

**Figure 6**: **(a)** Evolution of the experimentally determined fluorine concentration ($x_{act}$) as a function of the nominal fluorine concentration ($x$) for $PrFeAsO_{1-x}F_x$ samples prepared by the CSP-AP and HP-HTS methods. **(b)** Superconducting transition temperatures determined from magnetic susceptibility ($T_c^{mag}$) and electrical resistivity ($T_c^{onset}$) measurements as a function of the nominal fluorine concentration ($x$) for samples prepared by the CSP-AP and HP-HTS methods. The actual Fluorine content ($x_{act}$) was obtained by EDX compositional analysis and compared with the nominal fluorine concentration used during sample preparation.

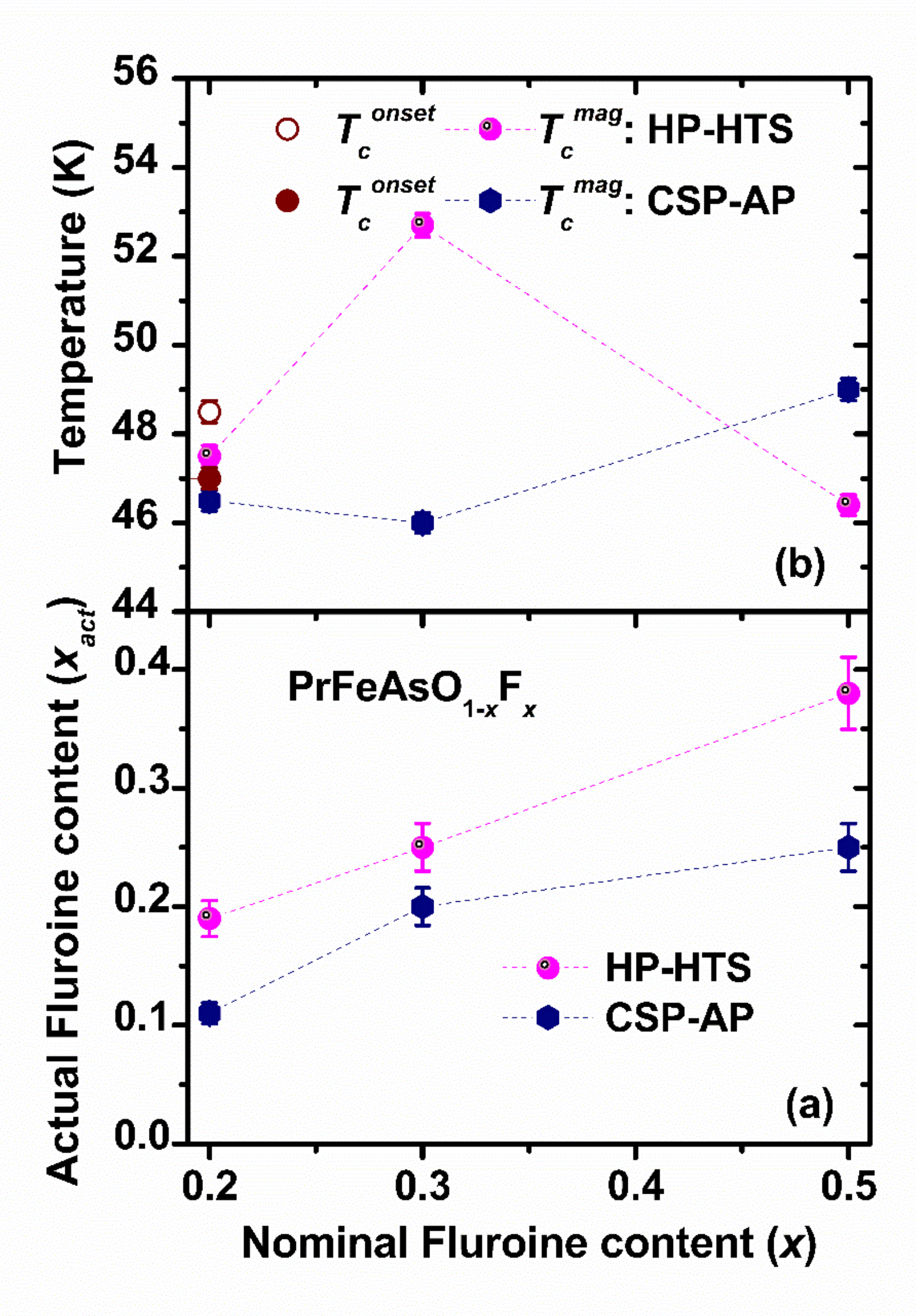